\documentclass[]{spie}  %>>> use for US letter paper
\usepackage{amsmath,amsfonts,amssymb}
\usepackage{graphicx}
\usepackage[colorlinks=true, allcolors=blue]{hyperref}
\usepackage{float}
\usepackage{gensymb}
\usepackage{wrapfig}

\title{The Space Coronagraph Optical Bench (SCoOB): 1. Design and Assembly of a Vacuum-compatible Coronagraph Testbed for Spaceborne High-Contrast Imaging Technology}

\author[a]{Jaren N. Ashcraft}
\author[a,c]{Heejoo Choi}
\author[b]{Ewan S. Douglas}
\author[a]{Kevin Derby}
\author[b]{Kyle Van Gorkom}
\author[a,b,c]{Daewook Kim}

\author[b]{Ramya Anche}
\author[a]{Alex Carter}
\author[b]{Olivier Durney}
\author[b]{Sebastiaan Haffert}
\author[b]{Lori Harrison}
\author[a]{Maggie Kautz}
\author[a]{Jennifer Lumbres}
\author[b]{Jared R. Males}
\author[a]{Kian Milani}
\author[b]{Oscar M. Montoya}
\author[a]{George A. Smith}

\affil[a]{James C. Wyant College of Optical Sciences, University of Arizona, Meinel Building 1630 E. University Blvd., Tucson, AZ. 85721}
\affil[b]{Department of Astronomy and Steward Observatory, University of Arizona, 933 N. Cherry Ave., Tucson, AZ 85719, USA}
\affil[c]{Large Binocular Telescope Observatory, University Of Arizona, 933 N. Cherry Ave. Tucson, AZ 85721} 
\authorinfo{Further author information: (Send correspondence to J.N.A. or E.S.D)\\J.N.A..: E-mail: jashcraft@email.arizona.edu,douglase@arizona.edu}

\begin{document}
\maketitle

% Ask Ewan about level of thermal control
\begin{abstract}
The development of spaceborne coronagraphic technology is of paramount importance to the detection of habitable exoplanets in visible light. 
In space, coronagraphs are able to bypass the limitations imposed by the atmosphere to reach deeper contrasts and detect faint companions close to their host star. 
To effectively test this technology in a flight-like environment, a high-contrast imaging testbed must be designed for operation in a thermal vacuum (TVAC) chamber.
A TVAC-compatible high-contrast imaging testbed is undergoing development at the University of Arizona inspired by a previous mission concept: The Coronagraphic Debris and Exoplanet Exploring Payload (CDEEP).
The testbed currently operates at visible wavelengths and features a Boston Micromachines Kilo-C DM for wavefront control.
Both a vector vortex coronagraph and a knife-edge Lyot coronagraph operating mode are under test.
The optics will be mounted to a 1 x 2 meter pneumatically isolated optical bench designed to operate at $10^{-8}$ torr and achieve raw contrasts of $10^{-8}$ or better.
The validation of our optical surface quality, alignment procedure, and first light results are presented. We also report on the status of the testbed's integration in the vaccum chamber.

\end{abstract}


% \tableofcontents

\section{Introduction}
    
    The Astro 2020 Decadal Survey has recommended the development of a future large infrared/optical/ultraviolet telescope optimized for observing habitable exoplanets as the highest priority for space frontier missions\cite{Astro2020}. Such a telescope would need to be equipped with high-contrast imaging technology capable of achieving deep contrasts of $10^{-10}$ and be capable of doing spectroscopy on the detected signal. 
    Spaceborne high-contrast imaging instruments are necessary to achieve this goal because they can entirely bypass the limitations on wavefront stability and transmission imposed by the atmosphere. The atmosphere is a fundamentally turbulent environment, so a separate adaptive optics (AO) system is necessary just to correct for the wavefront error that results from fluctuations in the optical path. These AO systems typically require one or more deformable mirrors just to accommodate the atmospheric aberrations\cite{Males2018}. The atmosphere also attenuates heavily near the ultraviolet, eliminating an entire region of spectral observation. 
    High-contrast imaging in space began with the Hubble Space Telescope (HST) and its repertoire of coronagraphic instrumentation\cite{NICMOS,ACS, STIS}. The HST continues to be an integral tool for astrophysical discovery, producing the data for the largest number of refereed papers in astronomy \cite{Crabtree2016} for a single observatory. Recently the James Webb Space Telescope (JWST) has launched, adding another promising space observatory equipped with high-contrast imaging instrumentation\cite{Krist2010,Boccaletti_2015} to the list of available science instruments.
    To aid in the pursuit of developing a future space observatory capable of $10^{-10}$ contrast, centers for testing technology and wavefront control methods are a necessity. Platforms for testing new technologies will invariably aid the pursuit of directly imaging faint astrophysical targets at high contrasts (e.g. exoplanets, debris disks).     
    Several high-contrast imaging testbeds exist among research institutions and universities around the world\cite{CAOTIC}. However, few are designed to operate at vacuum, most notably the testbeds at NASA's Jet Propulsion Laboratory (Decadal Survey Testbed\cite{Ruane2019}, HCIT\cite{Seo2019}) are presently demonstrating wavefront sensing and control in vacuum. 
    
    % Consider taking out the ones used solely for ground observation?
    % Source: https://sites.google.com/view/highcontrastlabs/testbeds?authuser=0
    % \begin{table}[H]
    %     \centering
    %     \begin{tabular}{|c|c|c|c|} 
    %         \hline
    %         Name & Institution & Vacuum ?  \\
    %         \hline
    %         Decadal Survey Testbed & NASA JPL & \checkmark \\
    %         HCIT & NASA JPL & \checkmark \\
    %         HCIL & Princeton University & X \\
    %         HCST & Caltech & X \\
    %         HiCAT & STScI & X\\
    %         % HOT & ESO & X\\
    %         JOST & STScI & X\\
    %         % LOOPS & Laboratoire d'Astrophysique de Marseille & X\\
    %         % MITHIC & Laboratoire d'Astrophysique de Marseille & X\\
    %         NEW EARTH Lab & National Research Council of Canada & X \\
    %         SAINT & NASA GSFC & X\\
    %         SPEED & Laboratoire Lagrange & X \\
    %         THD2 & Observatoire de Paris & X \\
    %         VODCA & University of Liege & X \\
    %         \hline
    %     \end{tabular}
    %     \caption{Table of high-contrast imaging testbeds meant to prototype spaceborne coronagraphic technologies\cite{CAOTIC}.}
    %     \label{tab:my_label}
    % \end{table}

	A new testbed for testing very high-contrast wavefront sensing and coronagraph technology in vacuum, the Space Coronagraph Optical Bench (SCoOB) is being developed by the University of Arizona's Space Astrophysics Laboratory, in close collaboration with the Center for Astronomical Adaptive Optics\footnote{https://www.as.arizona.edu/CAAO/}, the Large Optics Fabrication and Test Lab\footnote{https://www.loft.optics.arizona.edu/}, and the Extreme Wavefront Control lab\footnote{https://xwcl.science/}.
    To meet the challenge of imaging rocky exoplanets in reflected light, SCoOB will provide support to ongoing experiments in vacuum to prototype new high-contrast imaging systems.
    In this proceedings we report on the updates to the design of the testbed previously intended to prototype the Coronagraphic Debris and Exoplanet Exploring Pioneer (CDEEP) payload. 

\section{Review of Design and Assembly}

    The Space Coronagraph Optical Bench  is a vacuum-compatible high-contrast imaging testbed optimized for maximizing contrast on observatories with un-obscured pupils in small volumes. The testbench is meant to be a remotely-accessible high-contrast imaging laboratory that is uniquely suited to experiments in vacuum. Two coronagraph modes are currently undergoing modeling and development, a Vector-Vortex Coronagraph (VVC) and a Knife-edge Lyot Coronagraph (KLC). SCoOB's wavefront sensing and control algorithms are operated by the open-source CACAO platform\cite{Guyon2018}. This control software is used in other high-contrast imaging instruments (MagAO-X\cite{Males2018}, SCExAO\cite{Guyon2018}). Consequently, algorithms developed for these testbeds are easily migratable to SCoOB, and vice-versa.
    
    The baseline VVC mode was chosen for its achromaticity, high throughput, and ability to achieve small inner working angles (IWA) \cite{Mawet_2009}. This coronagraph was also the device chosen for the HabEx mission concept, a candidate next-generation observatory for the Astro 2020 Decadal Survey\cite{Astro2020}. The KLC mode was a simple addition requiring relatively few specialized components to operate, so it was also used during the initial development phase of the coronagraph as a test for our wavefront sensing and control algorithms.

    \subsection{Changes to CDEEP Coronagraph Design}
    
    The SCoOB testbed concept was previously a laboratory realization of a small space telescope coronagraph designed to be compact and entirely vacuum-compatible. Changes were made to the nominal design of the space coronagraph to realize the instrument at a lower cost based on the existing parent parabolas that our vendor (Nu-Tek) had in stock \cite{Maier_2020}. The changes from the OAP's used in the CDEEP design to the in-stock parabolas from Nu-tek are shown below in Tables \ref{tab:opticsroc} and \ref{tab:opticsoad}.
     
    \begin{table}[H]
        \centering
        \begin{tabular}{|c|c|c|c|}
        	\hline
             & CDEEP Design & SCoOB Design & Difference  \\
             \hline
            OAP 0  & -142 & -293.6 & 151.6 \\ 
            OAP 1 & -200 & -254 & 54\\
            OAP 2 & -202 & -346 & 144\\
            OAP3 & -900 & -914.4 & 14.4\\
             \hline
        \end{tabular}
        \caption{Change in radius of curvature from the nominal CDEEP design}
        \label{tab:opticsroc}
    \end{table}

	\begin{table}[H]
	\centering
	\begin{tabular}{|c|c|c|c|}
		\hline
		& CDEEP Design & SCoOB Design & Difference  \\
		\hline
		OAP 0  & 34 & 27 & 7 \\ 
		OAP 1 & 72 & 40 & 32\\
		OAP 2 & 73 & 55 & 18 \\
		OAP3 & 134 & 100 & 34\\
		\hline
	\end{tabular}
	\caption{Change in off-axis distance from the nominal CDEEP design}
	\label{tab:opticsoad}
\end{table}
    
    The final set of optics are more on-axis, and the design is unfolded so that room could be made for commercially-available off-the-shelf mounting hardware. The result is a testbed that is less prone to misalignment and polarization induced wavefront errors, increasing the potential performance beyond what was baselined for the original CDEEP concept\cite{Maier_2020}. 
    
    \begin{figure}[H]
        \centering
        \includegraphics[width=\textwidth]{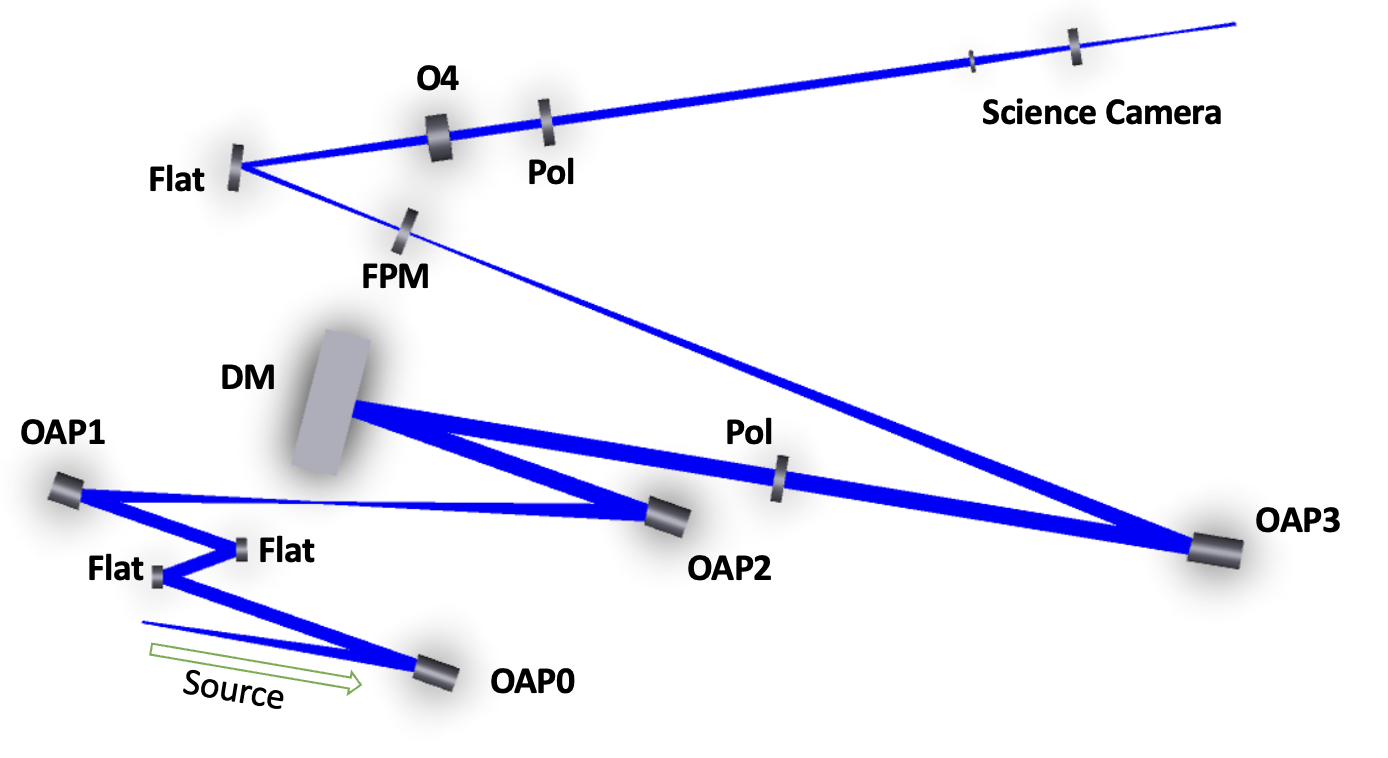}
        \caption{Raytrace model of SCoOB in Zemax OpticStudio. The optical path begins on the bottom left where a source simulator generates an isotropic point-source before OAP0. Light then propagates through the coronagraph into the science camera.}
        \label{fig:testbed_raytrace}
    \end{figure}

	The optical path (shown in figure \ref{fig:testbed_raytrace}) begins with a source simulator. Most experiments were conducted using a point-source microscope to illuminate a small (15um diameter) pinhole at this location, but recently this was switched to a single-mode fiber tip with a sufficiently small mode field diameter ($<5um$) to act as a point source. The light from the source is collimated by OAP0 which propagates to the first pupil position at the first flat mirror. We plan to replace this flat with a piezo-electric fast-steering mirror for fine steering control and telescope pointing tests. A mirror tilted the same angle in the opposite direction follows to not induce polarization aberrations in the testbed and maintain a small footprint. Two OAPs follow this optic (OAP1, OAP2) which relay the image of the pupil onto a Boston Micromachines Kilo-C 1.5um stroke DM. After the DM, the light passes through a circular polarizer and then on to OAP3. OAP3 has the largest focal length of all optics in this system. This is to deliver a beam of appropriate focal ratio (F/48) to the focal plane mask. After the focal plane mask a fold mirror is inserted to redirect the beam to a collimating lens (O4) which passes the beam onto the next circular polarizer and then Lyot Stop. A lens group follows the Lyot stop to form an image on the AWO science camera.
	
	The flat after the focal plane mask was selected in order to ensure that the testbed was sufficiently compact. Originally this optic was an off-axis parabola, but due to the space constraints on the testbed and the focal ratio constraint on the beam incident on the focal plane mask, the exit pupil was 1mm in radius. The corresponding Fresnel diffraction effects added amplitude variation across the pupil, which we believed could limit our contrast. The OAP was replaced with a flat and a collimating lens of a longer focal length to increase the exit pupil size. A diverging beam incident on a flat surface typically induces some spherical aberration. However, given the slow focal ratio of the beam the aberration induced is entirely negligible. We also don't expect this change to contribute substantive polarization aberrations that would corrupt our contrast because it is after the focal plane mask.
	
	\subsection{Mount Designs and Tolerance Analysis Results}
	
	High-contrast imaging instruments require a well-characterized optical design in order to achieve diffraction-limited optical imaging. A diffraction model informed tolerance analysis was conducted in order to asses the sensitivity of the coronagraph to low-order aberrations. Using HCIPy\cite{por2018hcipy} we propagated wavefront error expressed with Zernike polynomials through a VVC and examined the average contrast in the dark hole region for various applied amplitudes of individual Zernike modes.  A separate tolerance analysis in Zemax OpticStudio was done to examine the Zernike-decomposed wavefront error induced by perturbations of the optics in position and angle. Using these two analyses in tandem allowed us to examine the degradation in coronagraph contrast in response to mechanical aberration.  The results of the tolerance analysis are shown in in Figure 6 and 7 of Maier et al\cite{Maier_2020}, but a table of the final tolerances for the optics used in SCoOB are shown in Table \ref{tab:tolerances} below.
	
	\begin{table}[H]
		\centering
		\begin{tabular}{c c c c c c c c c c c c c}
			\hline
			& & OAP0 & & & OAP1 & & & OAP2 & & & OAP3 \\
			& x & y & z & x & y & z & x & y & z & x & y & z \\
			\hline
			Decenter [mm] & 0.1 & 0.1 & 0.3 & 0.1 & 0.1 & 0.1 & 0.1 & 0.1 & 0.1 & 0.1 & 0.1 & 0.1  \\
			Tilt [deg] & 0.1 & 0.1 & 0.3 & 0.02 & 0.02 & 0.1 & 0.05 & 0.3 & 0.5 & 0.2 & 0.3 & 0.5  \\
			\hline
			& 
		\end{tabular}
		\caption{Sensitivities of each optic in 6 degrees of freedom such that the RSS of the contributions from individual Zernike modes do not degrade contrast beyond 10$^{-8}$.}
		\label{tab:tolerances}
	\end{table}
	
	To isolate the optics from any residual vibration, the mounts are seated on 1" passivated posts from Newport. The mounts for the OAPs are vacuum-compatible variants of the THORLABS K6XS-SM1 6-axis kinematic mounts. A quarter-turn of each adjustment knob was defined as the "sensitivity" of the degree of freedom. The K6XS-SM1 is sufficiently sensitive to achieve the tolerances outlined in table \ref{tab:tolerances}. 

	To best match the CDEEP space coronagraph design, the clear aperture of the off-axis parabolas were smaller than are suitable to most commercially available optical mounts (1", 1/2", etc.). To adapt the mirrors to commercial off-the-shelf optomechanics, interface cups were designed to mount the mirrors in. The cups were designed such that the inner diameter of the mounting cylinder matched the outer diameter of the OAP. The OAP was bonded to the cup with 2216 epoxy adhesive and allowed to cure with a cylindrical shim in place to keep the optic centered during the curing. After the curing was completed the shim was removed and the OAP in the cup was mounted to the THORLABS K6XS mounts. 
	
	\begin{figure}[H]
		\centering
		\includegraphics[width=0.5\textwidth]{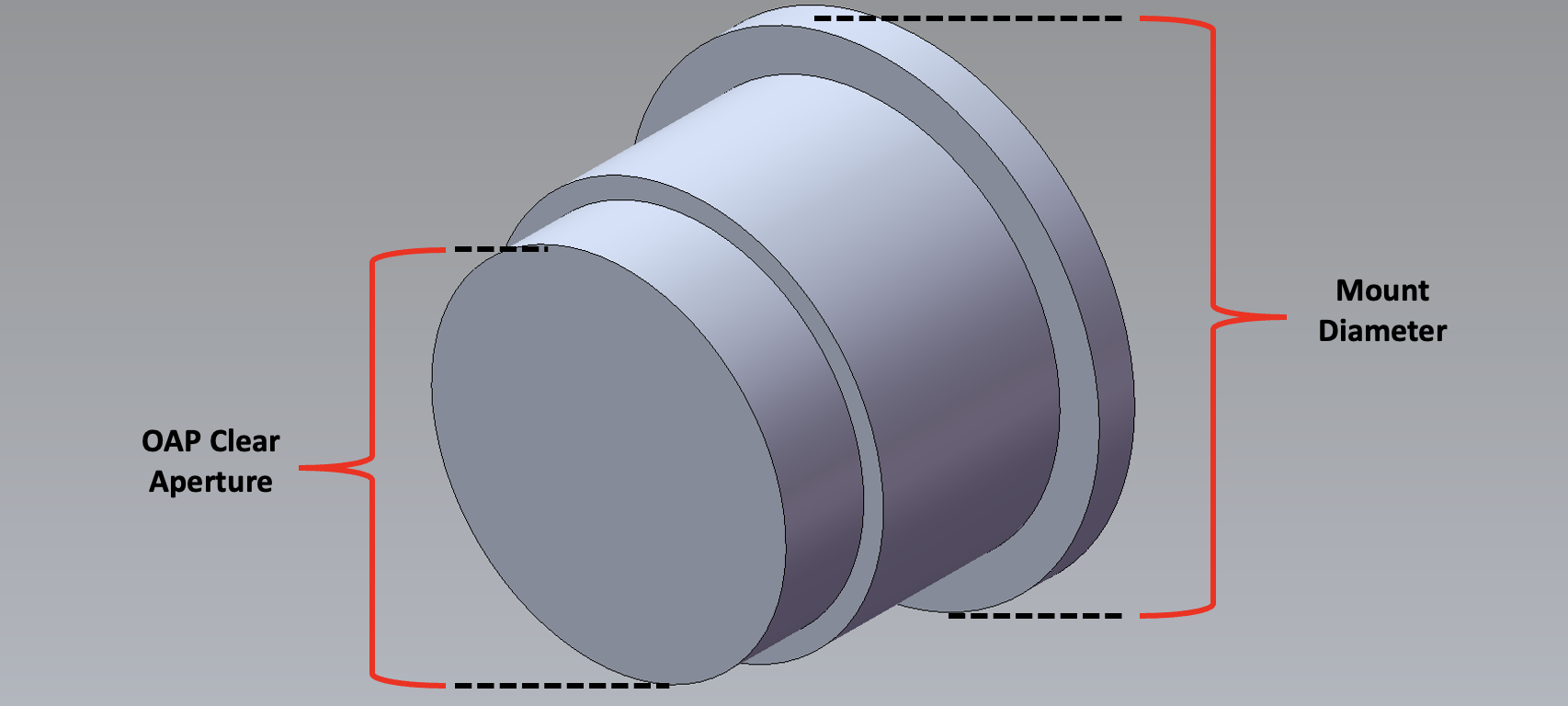}
		\caption{Solid model of the interface cup holding an off-axis parabola. These cups make optics of any size under the mounting diameter adaptable to commercially available optomechanics at relatively low cost.}
		\label{fig:oap_cups}
	\end{figure}

	The optics are atypical in aspect ratio with the length of the OAP being larger than the clear aperture. This was purely a consequence of using existing parabola parents from Nu-Tek, but we expect less mount-induced aberration because of it. Trefoil is a common aberration induced by 3-point optical mounts which result from the deformation of the surface by the mounting hardware. Because of the skewed aspect ratio and the interface cups, we expect our optics to be more resilient to surface deformation induced by the optomechanics.
	
	\subsection{OAP \& Flat Measurement}
	
	Upon assembly of the optics into their associated mounting hardware, a preliminary characterization run was conducted to verify the quality of the OAPs. A simple and effective test of the surface quality is done interferometrically. We used a 4D PhaseCam 6000 Interferometer in a double-pass configuration. The OAP under test is placed in the path of the beam exiting the interferometer. After passing the OAP the beam is incident onto a reference optic which retroreflects the beam back through the system. We used the PhaseCam with a focusing optic to create a simulated source that the OAP could collimate onto a reference flat. The flat mirror only has two degrees of freedom (tip, tilt) because it needs to be normal to the incoming beam to retroreflect. This reduces the number of degrees of freedom that needed to be aligned, which simplifies the metrology process. The remaining complication was to position the OAP in the correct orientation, the procedure for which is outlined below. The theory of measuring the optics relies on aligning each degree of freedom individually, and iteratively repeating the process until a satisfactory interferogram is achieved. The coordinate system definitions for the following procedure are such that the z-axis is the initial propagation axis, the y-axis is in the direction transverse to the optical bench, and the x-axis is orthogonal to both. Before alignment of the OAP to the interferometer, we define the optical axis by placing two parallel irises on an optical rail in front of the interferometer. Translation stages under the interferometer were used to finely tune its position such that light exiting the interferometer passed through both irises. Upon sucessful definition of an optical axis, the position of the interferometer was fixed and the OAP could be aligned to it using the following procedure:
	
	\begin{enumerate}
		\item \emph{Clocking / Z-Tilt}: Have the OAP vendor mark the direction of the parent axis of the OAP under test. Then, clock the OAP until the marker is aligned to the horizontal on the 6DOF mount. This degree of freedom tends to not be as sensitive to misalignment (see Table \ref{tab:tolerances}), so this method should be sufficient. 
		\item \emph{X-Decenter}: Place a reference target (e.g. a card) below the image produced by the interferometer. Then place the OAP on a rail in front of this target, confining the interferometer and OAP to the Y-Z plane. 
		\item \emph{Y-Decenter}: Translate the OAP vertically such that the beam is approximately centered on the optical axis defined by the interferometer and two irises.
		\item \emph{Z-Decenter}: Translate the OAP along the rail a distance equal to its focal distance with respect to the target placed in step 2. This step is made easier using a rail with regular rulings, and can be refined with the adjustment knobs on the 6DOF mount later.
		\item \emph{Y-Tilt}: Tilt the OAP about its Y-axis by the off-axis angle of the OAP. This can be accomplished coarsely with a protractor as a visual aid, and can be refined interferometrically with the tilt knob on the 6DOF mount.
		\item \emph{X-Tilt}: Adjust the X-tilt knob until the beam after the OAP is approximately level with the X-Z plane.
		\item Place the reference flat in the path of the ensuing quasi-collimated beam. Adjust the flat in tip and tilt until the beam is visible on the detector and no straight-line fringes are observed.
		\item  If there are a large amount of remaining focus fringes, carefully adjust the OAP's position on the rail in small increments. If there is a small amount of focus fringes remaining, adjust the tip/tilt/z knobs in equal amounts until the focus fringe is nulled.
	\end{enumerate}

\begin{figure}[H]
		\centering
		\includegraphics[width=0.7\textwidth]{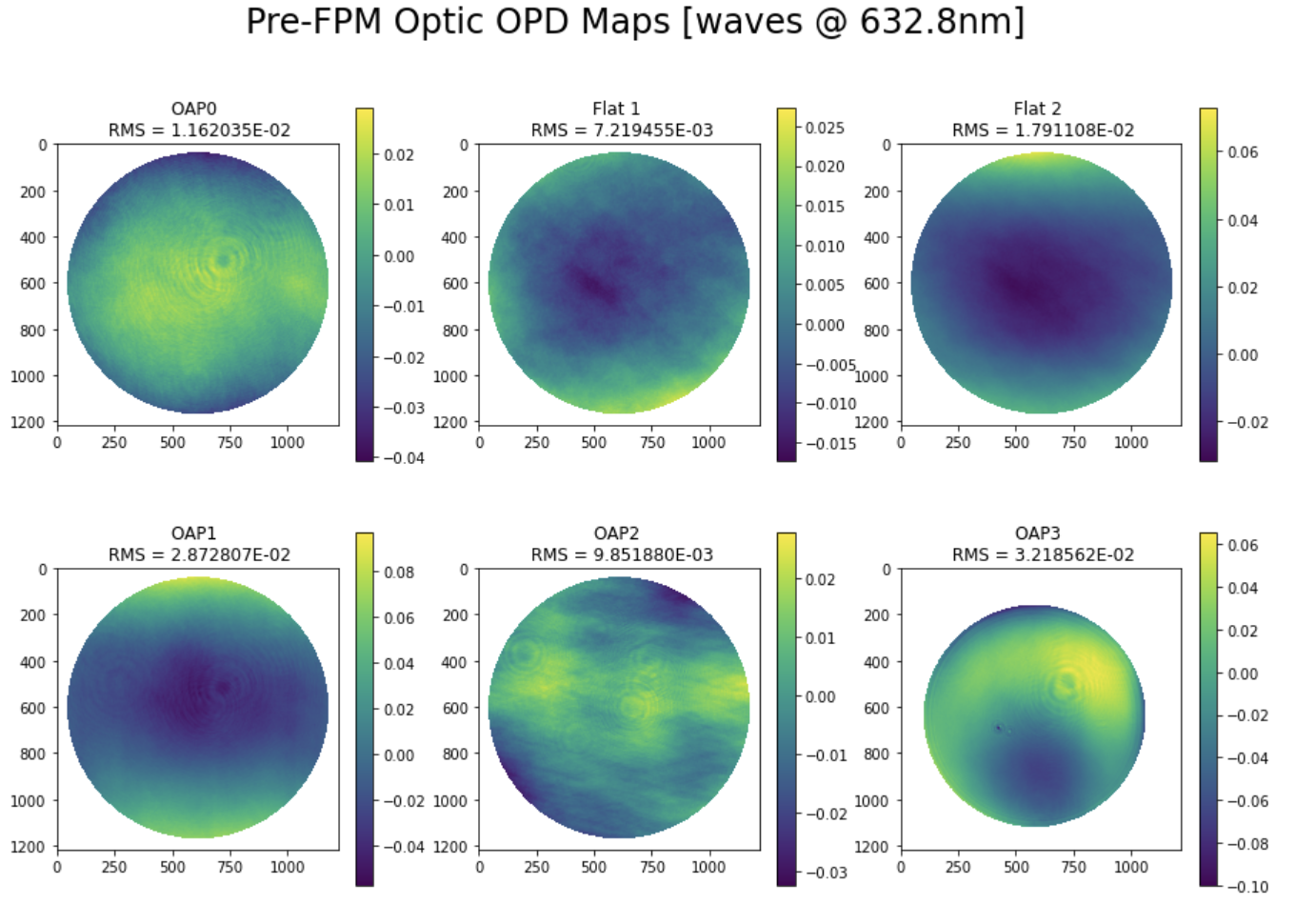}
		\caption{Figure with the OPD of the testbed optics shown in units of waves tested at 632.8nm. On the top row from left to right is OAP0, the first, and second flat mirrors. The bottom row is OAP1, OAP2, and OAP3. Only the optics upstream of the focal plane mask were tested because they determine the shape of the PSF delivered to our coronagraphic mask, which sets our system's performance. OAP3 is slightly over-filled by the 4D beam due to the long focal length, so the OPD map is slightly smaller than the rest. There is some residual alignment-induced aberration that was challenging to null in OAP3 due to its sensitivity to tilt, but the measured wavefront error was below our specifications so we proceeded with assembly.}
		\label{fig:optic_opd}
	\end{figure}

	This procedure will reveal what aberrations persist through the systems, so the remaining steps will largely depend on the interferogram observed. Dealing with residual coma and astigmatism is nontrivial because the aberrations will typically coupled to multiple degrees of freedom. They are indicative of primarily X- and Y-tilt errors, so adjusting these on the OAP mount and then accommodating for the adjustment with the tip/tilt knobs on the reference flat can slowly eliminate the aberration. If the aberration amplitude is too large, simply repeat the procedure above and then begin refining nulled interferogram until a satisfactory result is achieved. The results of our interferometric testing using this procedure are shown in Figure \ref{fig:optic_opd}. 
	
\subsection{Alignment plan, Assembly, and performance validation tests}

The philosophy for aligning the testbed optics is similar to the interferometric testing of the OAPs. Each degree of freedom is adjusted at a coarse and fine level, and then revisited until the interferogram is sufficiently nulled. We take the plane of the testbed to be the x-z plane, and the axis transverse to the testbed to be the y axis. All DOF are considered local to the optical element (e.g. z is a tilt about the axis through the center of an OAP).

\begin{enumerate}
	
	\item Place OAP’s with 6DOF mounts on a 4” post, this defines the nominal height of all optics in the testbed.
	\item Place a 1” iris on a post and then in a post holder. Adjust the y-decenter of the iris to be coaxial with the OAPs. This will be our rough y-decenter reference for the testbed.
	\item Use digital calipers to set the clear diameter of the iris to be equal to the beam size at the deformable mirror (9.2mm)
	\item Position a point-source microscope in a mount with adjustable height (translation stage, or stacked posts + spacers) such that the beam is coaxial to the iris.
	\item Place the pinhole assembly on the testbed breadboard to define the first point of the optical axis.
	\item Place the source with a focusing optic behind the pinhole and adjust the position until there is sufficient transmission through the pinhole. We used a point-source microscope for our initial source, which has an internal camera that is conjugate to the source focus, significantly reducing the pinhole alignment difficulty. After sufficient throughput is achieved the pinhole must remain fixed.
	\item Place OAP0 one focal distance away from the pinhole plane using a ruler, and adjust the position until the beam after the mirror is approximately collimated when viewed on a card. Then fix the position of OAP0 to the testbed using a clamping fork. 
	\item Place a shear plate in the ensuing quasi-collimated space normal to the incoming beam and observe the interferogram. If the fringes are not parallel to the marked line, then there is residual focus in the beam.
	\item Rotate the shear plate by 90 degrees by removing it from the post and placing it back in. If the fringes are tilted at a different angle than the previous configuration, there is astigmatism in the beam. Use the tip/tilt/z knobs to translate out the residual focus error shown in the shear plate. The astigmatism is addressed by altering the tilt angle about the X and Y axis to null the observed fringes, rotating the shear plate, and repeating the process.
	\item Place a reference flat in the collimated space to retroreflect the beam back into the point-source microscope to view the PSF in detail. Examining the shape of the PSF is indicative of the remaining aberrations present in the system. This will serve as a guide for what DOF require fine-tuning before proceeding with the next optic.
\end{enumerate}

This procedure is sufficient for collimating a beam from the source simulator and ensuring that minimal additional aberration is introduced to the ensuing beam. To align an OAP to a collimated space, begin from step 7 in the procedure above and instead use the following procedure:

\begin{enumerate}
    \item Place OAP1 focal distance away from the intermediate pupil plane using a ruler, and adjust the position until the beam after the mirror forms a focus at approximately the correct distance when measured with a ruler. Then fix the position of OAP0 to the testbed using a clamping fork. 
    \item Place a card at the focal plane to assess the presence of any large aberrations that are easily nulled by a rotation about the y- or x- axes.  
    \item Once the point-spread function is smaller than is viewable by eye, place a detector at the focal plane.
    \item Use the tip and tilt knobs to minimize the aberration in the PSF. The dominant modes of aberration from tip and tilt are astigmatism and coma.
    \item Upon successful nulling a retroreflecting sphere can be placed in the image plane to view the PSF in the point-source microscope. Viewing the aberrations in double-pass enhances the small amplitude aberrations that are a result of decenter in x, y, and z. 
\end{enumerate}

Ultimately the fine alignment of the OAP's are somewhat exploratative because of how dependent the alignment is on the initial setup. We found success in approaching the alignment of the testbed by iteratively exploring one degree of freedom at a time to minimize the coupling of wavefront error from orthogonal degrees of freedom. The procedures listed above are repeated for OAP2 and OAP3. Upon successfully aligning the OAPs up to OAP3 we evaluated the as-aligned wavefront error at the coronagraphic focal plane by placing a ball bearing in OAP3's image plane. The resultant interferogram was viewed in double-pass to observe if our alignment efforts were successful. 

\begin{wrapfigure}{r}{0.5\textwidth}
	\centering
	\includegraphics[width=0.5\textwidth]{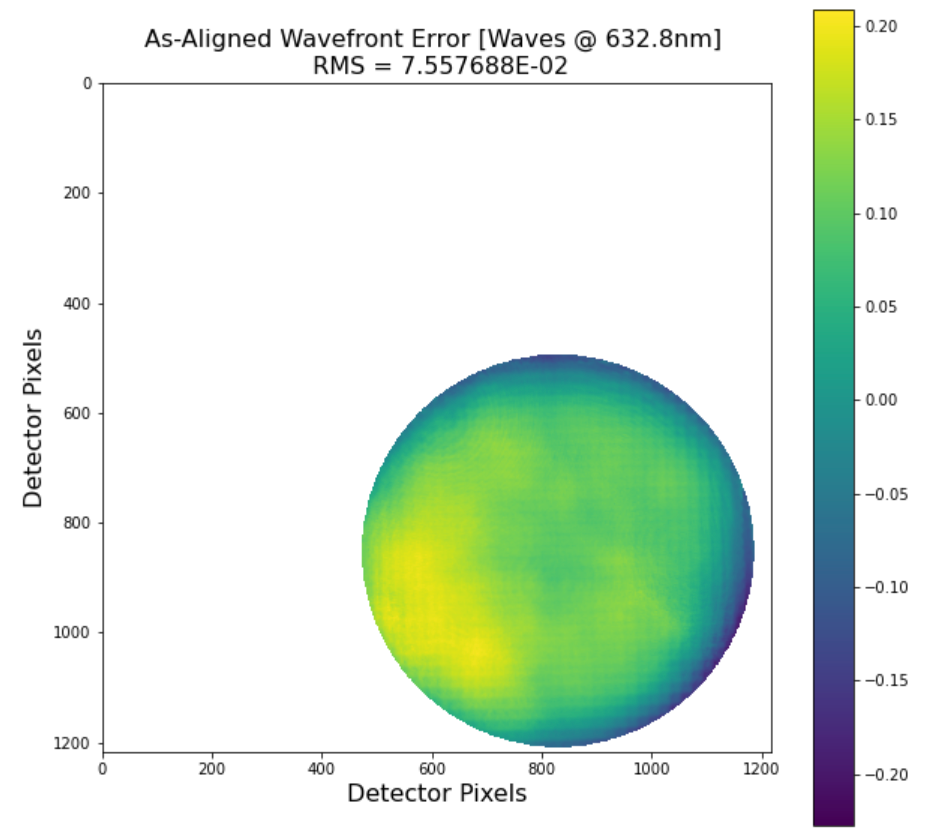}
	\caption{The as-aligned wavefront error map measured by the 4D interferometer in double pass. Units are in waves at the test wavelength of 632.8nm. The pupil of the system is smaller than that of the 4D interferometer so the interferogram occupied a region within the resultant interferogram file. On this scale, the spacing between actuators of the deformable mirror is visible.}
	\label{fig:as_aligned_wfe}
\end{wrapfigure}

At the test wavelength we achieved roughly $\lambda / 40$ RMS wavefront error in single-pass,  which was lower than our wavefront error specification outlined in Maier et al \cite{Maier_2020}. Given satisfactory optical alignment, we could proceed with the alignment of our spatial filters and begin coronagraphic imaging. Aligning to the knife-edge mask is fairly straightforward given that it should just block the core of the instrument PSF. Inserting the knife on a translation stage until a large amount of light is visibly reflected back is sufficient, and can be refined simply by observing the knife at the science camera. However, aligning to the VVC can be challenging because of its transparency. Luckily, the modern standard is that there is always a compact polarized screen somewhere. Aligning to the VVC is made simple by viewing it through a polarizer crossed with respect to the polarization of a screen (e.g. of a Mobile Phone). The result is a clear view of the optical vortex. Turning up the source power and/or opening the iris also allows you to view the light scattered off of the surface of the VVC from the PSF. Getting the center of the vortex with the spot of scattered light coarsely aligned will get the system mostly aligned. The fine allignment to the singularity can be achieved viewing the beam in a pupil plane downstream.

\begin{figure}[H]
	\centering
	\includegraphics[width=0.8\textwidth]{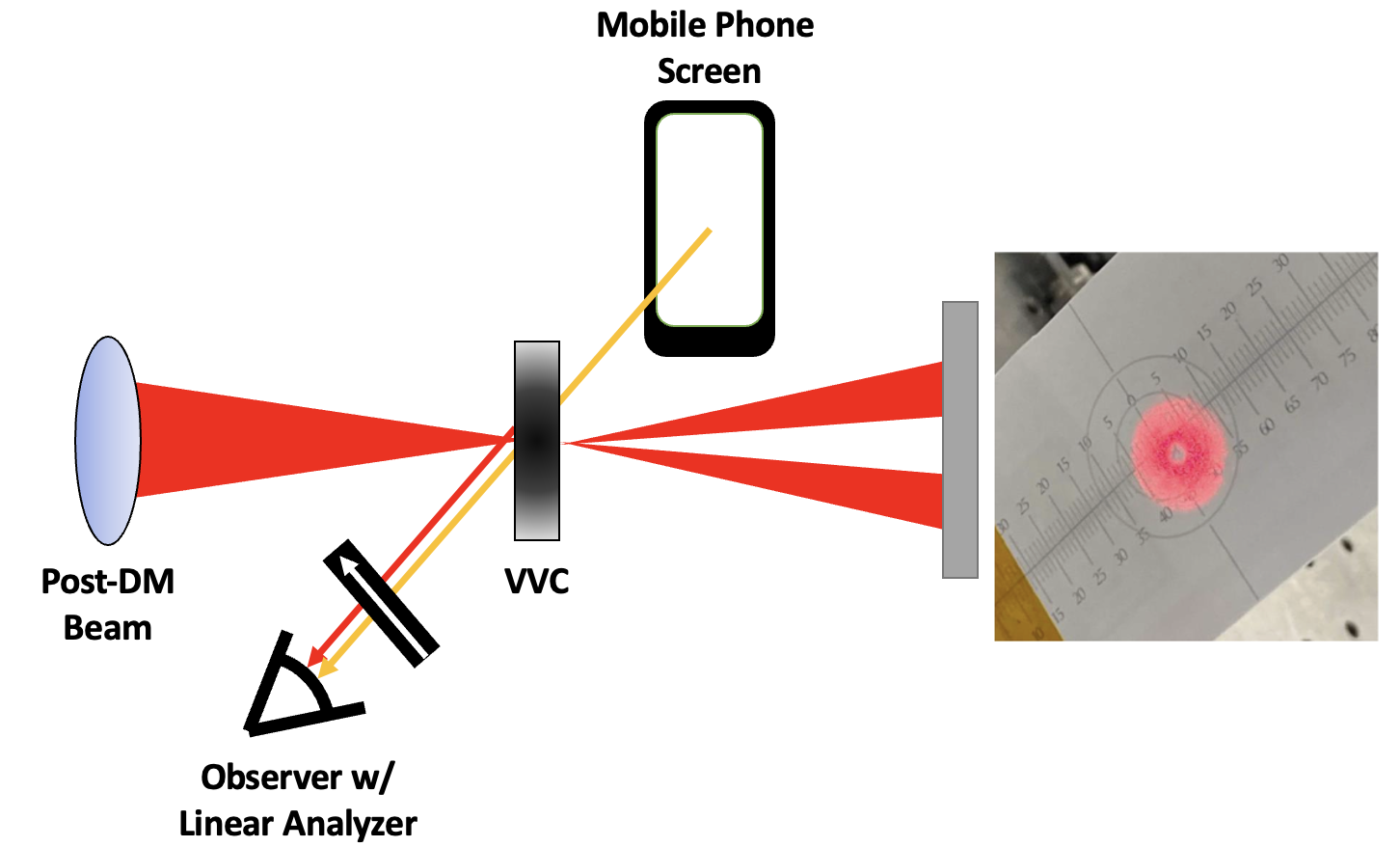}
	\caption{Diagram illustrating the fine alignment to a VVC using a polarizer and mobile phone screen. The transparency of the mask and small singularity can be very challenging to align to. In the presence of polarized illumination the vortex pattern is easily seen. Simultaneously viewing the laser light reflecting off the mask and the vortex pattern considerably eases the alignment to these masks, and only uses materials that would be readily available in any laboratory.}
	\label{fig:vvc_align}
\end{figure}

\begin{figure}[H]
	\centering
	\includegraphics[width=0.8\textwidth]{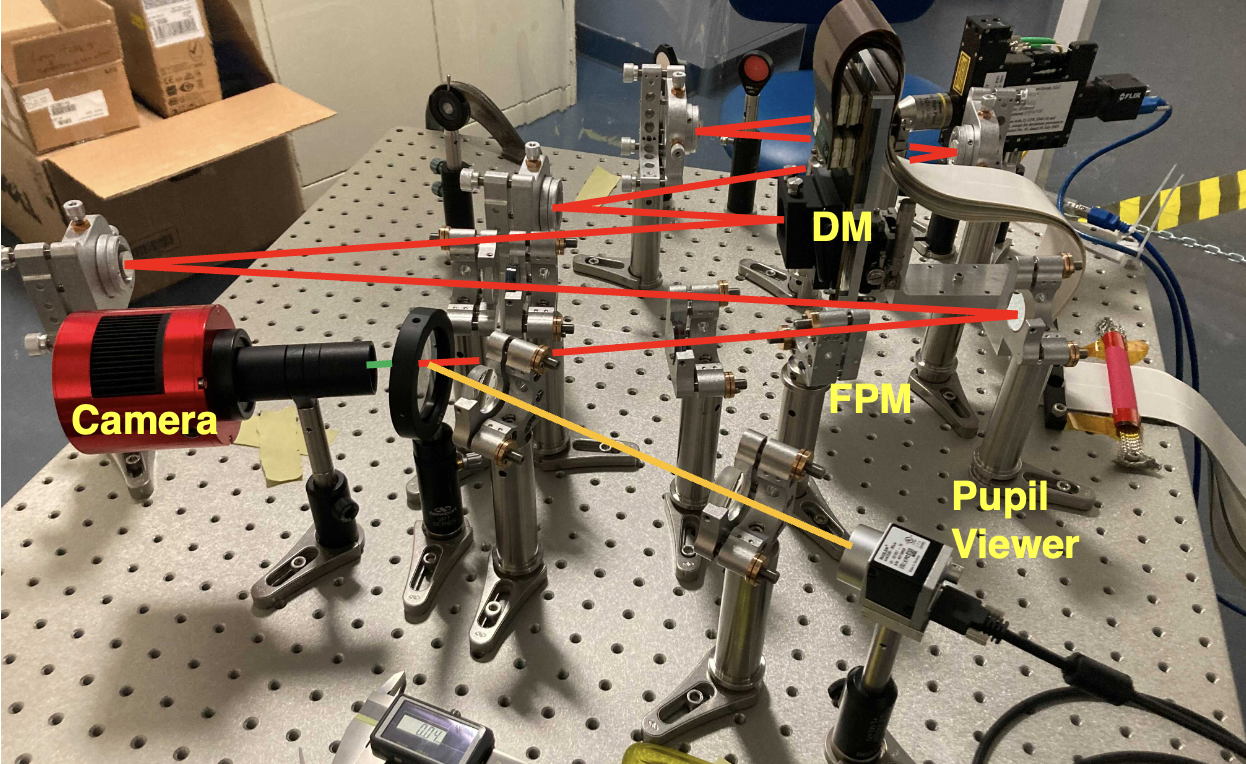}
	\label{fig:assembled_testbed}
	\caption{SCoOB Assembled in our laboratory's clean tent. }
\end{figure}

\phantom{} \\

\section{First Light Results}

Upon successful installment of the hardware in the testbed we examined the point-spread function and it's influence on the two coronagraph modes. Before observation, whatever residual low-order aberrations were cleaned up with the Eye Doctor\cite{Bailey_2014} algorithm developed for the Large Binocular Telescope. The eye doctor employs a grid search of low-order zernike aberrations applied to the DM to determine which coefficient maximizes the strehl ratio of the PSF. The summed shape of these aberration modes determines the "flat" position of the deformable mirror to further enhance our image quality. Below are figures of the testbed's point-spread function, and coronagraphic image plane.
    
    \begin{figure}[H]
        \centering
        \includegraphics[width=\textwidth]{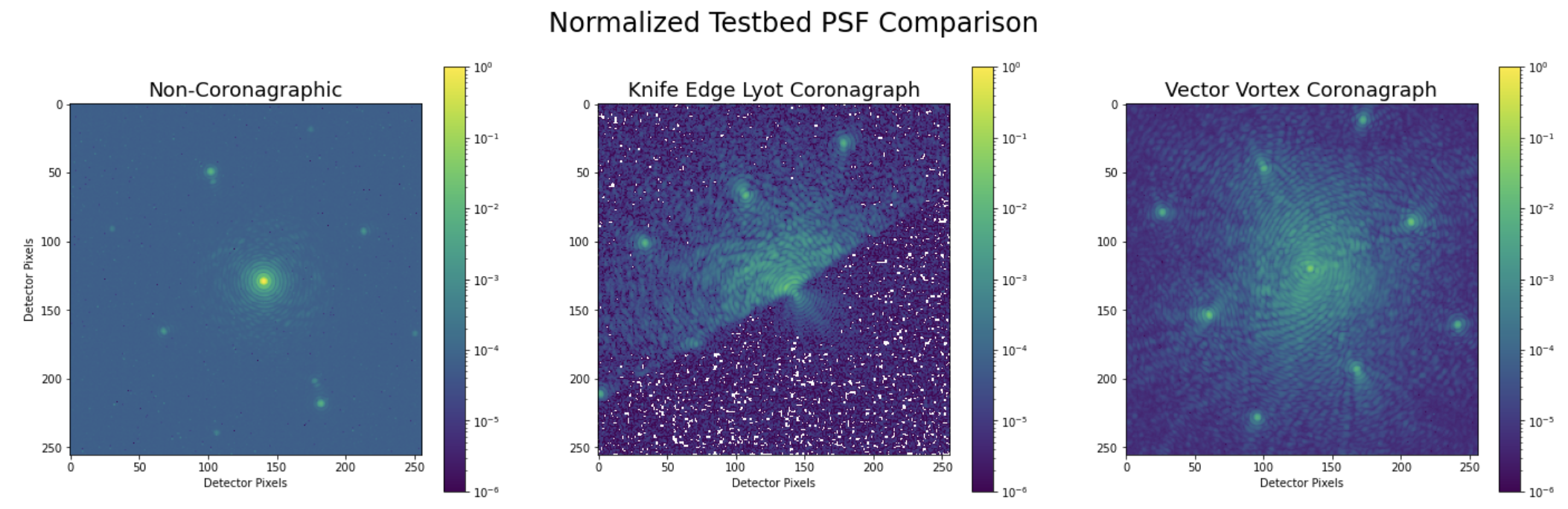}
        \caption{(Left) First light measurement of the testbed point-spread function normalized to the peak of the airy disk. (Middle) The knife-edge Lyot coronagraph mode image plane plotted on the same vertical scale as the non-coronagraphic PSF. (Right) The vector vortex coronagraph mode image plane illustrating the on-axis starlight rejection being comparable to the background. The fringes observed in this image result from the self-coherent camera pinhole in the Lyot stop.}
        \label{fig:psf_compare}
    \end{figure}
    
The coronagraphic mask and Lyot stop only serve to contribute part of the contrast. In order to reach deeper contrasts we employ wavefront sensing and control techniques suitable for high-contrast imaging. A more comprehensive review of the implementation and results of these efforts are outlined in Van Gorkom et al\cite{VanGorkom2022}, but a brief introduction to current efforts is written here.

\emph{Electric field conjugation} (EFC)\cite{Giveon:09} is a method of focal plane wavefront sensing that has seen widespread implementation among high-contrast imaging instruments used for astrophysics\cite{Kasdin_2020,Ahn2021}. The method relies on an accurate Fresnel/Angular Spectrum forward model to propagate the effects of phase errors on the optical surfaces to the image plane. This model-assisted approach is capable of reaching very deep contrast levels. In the small aberration regime, the relationship of the electric field in the pupil plane to that of the image plane is approximately linear. Using this relation solutions to the electric field of the image plane can be probed for a given control region. Upon succesful sensing of the electric field in this region, the opposite phase can be applied to the image field to create a high-contrast region (near $10^{-8}$ has been demonstrated in SCoOB). In practice, this method is limited by the dynamic evolution of the electric field as a function of time. Vibrations, ambient turbulence, and temperature differentials are all sufficient to aberrate the electric field such that the contrast is lost. However, the EFC algorithm is computationally time-consuming and cannot be easily recalculated. Therefore, an efficient method of maintaining the dark hole is necessary for proper operation. Methods of circumventing this limitation are being explored by other investigators using an algorithm called Implicit Electric Field Conjugation \cite{haffert2022magaox}. To see the results of this method, please consult our other manuscript published in these proceedings \cite{VanGorkom2022}.

\emph{The Self-Coherent Camera} (SCC)\cite{Janin_Potiron_2016,Derby22} is a method of focal-plane wavefront sensing undergoing experimentation on SCoOB that is uniquely suited to the dark hole maintenance problem. By placing a pinhole in the opaque part of the Lyot stop, the residual starlight that would have ordinarily been blocked is allowed to pass. Because stars are partially coherent with themselves\cite{goodman}, this create interference fringes with the starlight, but not the planet light. Transforming the signal into the fourier domain reveals that this effect creates side-bands in the Optical Transfer Function (OTF) where the wavefront phase can be extracted. This technique requires no forward model in order to do wavefront sensing, so it can be computed faster than EFC. Given a sufficient starting solution and a small pinhole, the SCC is capable of maintaining deep contrasts in the dark hole\cite{Derby22}.

\section{Vacuum Chamber}

SCoOB will be placed in an Rydberg Vacuum Sciences 104430 Thermal Vacuum (TVAC) Cycling chamber that has been acquired by the UA Space Astrophysics Laboratory. The chamber is in an underground room to minimize the coupling of external vibrations from the building. The TVAC chamber employs a dry screw pump and turbomolecular pump to achieve high vacuum. Upon achieving base pressure, temperature is regulated by a combination of nichrome heaters and gaseous nitrogen from boiled LN2. Thermal and vacuum cycling is conducted using a recipe-based system to simplify experimentation. We anticipate beginning preliminary in-chamber tests before the end of the year.

\begin{table}[H]
    \centering
    \begin{tabular}{|c|c|}
    \hline
    Specification & Value \\
    \hline
        Diameter & 1.2 m \\
        Length & 2.2 m  \\
        Interior Finish & $\# 4$ grained finish \\
        Roughing Pump & Varodry VD65 dry screw pump \\
        High-vacuum Pump & Mag-Lev Turbomolecular pump \\
        Temperature Range & -150 \degree C to 150 \degree C \\
        Thermal Stability - Room Temperature & $\pm$ 2 \degree C \\
        Thermal Stability - 150 \degree C & $\pm$ 10 \degree C \\
        Ramp Rate range & $\pm$ 1 \degree C/hour to $\pm$ 2 \degree C/min\\
        Design vacuum & 1e-8 torr
        \\
        \hline
    \end{tabular}
    \caption{Table of TVAC chamber operational specifications, some of which are currently being evaluated.}
    \label{tab:vaccumspecs}
\end{table}

\begin{figure}[H]
    \centering
    \includegraphics[width=0.75\textwidth]{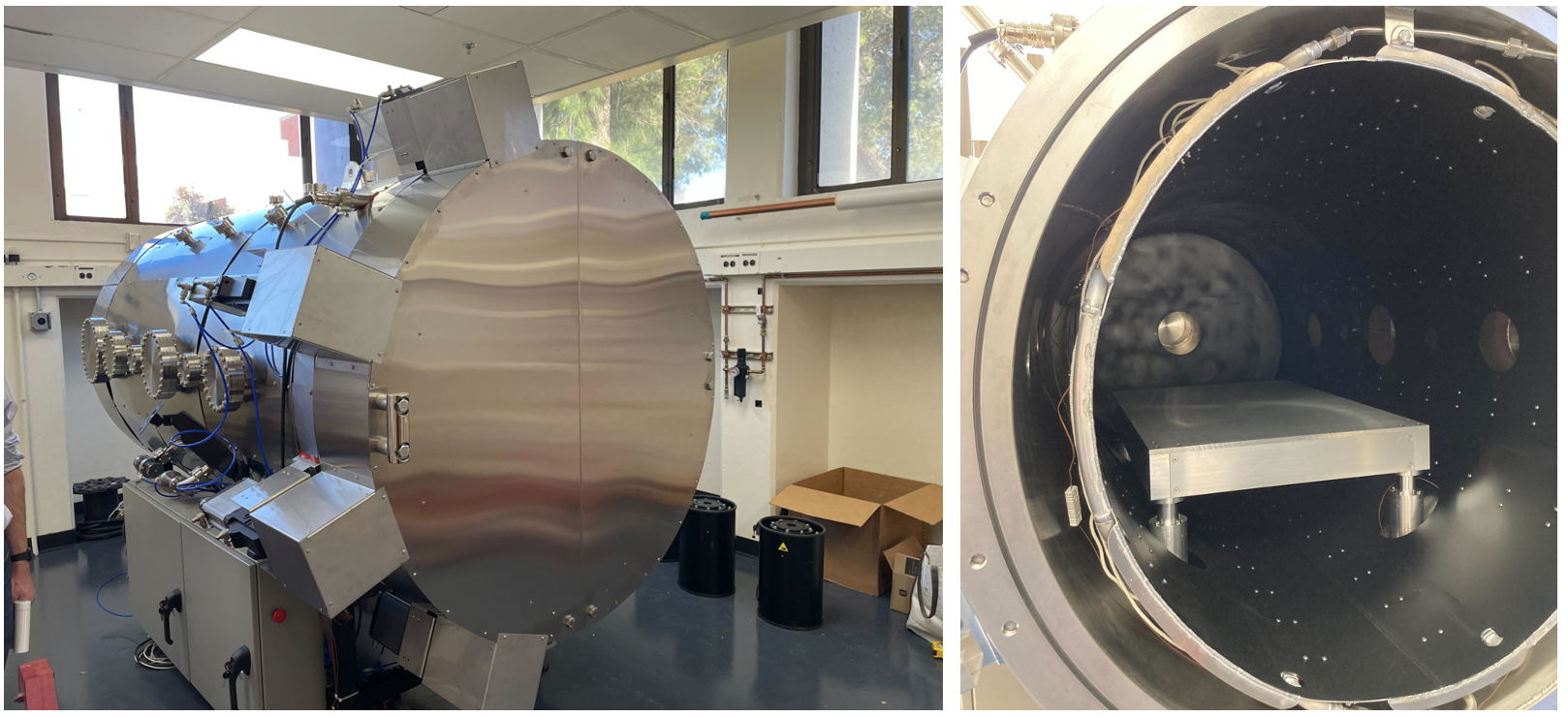}
    \caption{RVS 104430 TVAC chamber at the basement of Steward Observatory. (Left) The exterior of the cylindrical testbed chamber. (Right) The interior of the vacuum chamber showing the support table. A 4-inch Newport high-vacuum optical breadboard will be installed in the chamber to support the SCoOB breadboard.}
    \label{fig:tvac_chamber}
\end{figure}

\section{Conclusion and Future Work}

\begin{figure}[H]
    \centering
    \includegraphics[width=0.8\textwidth]{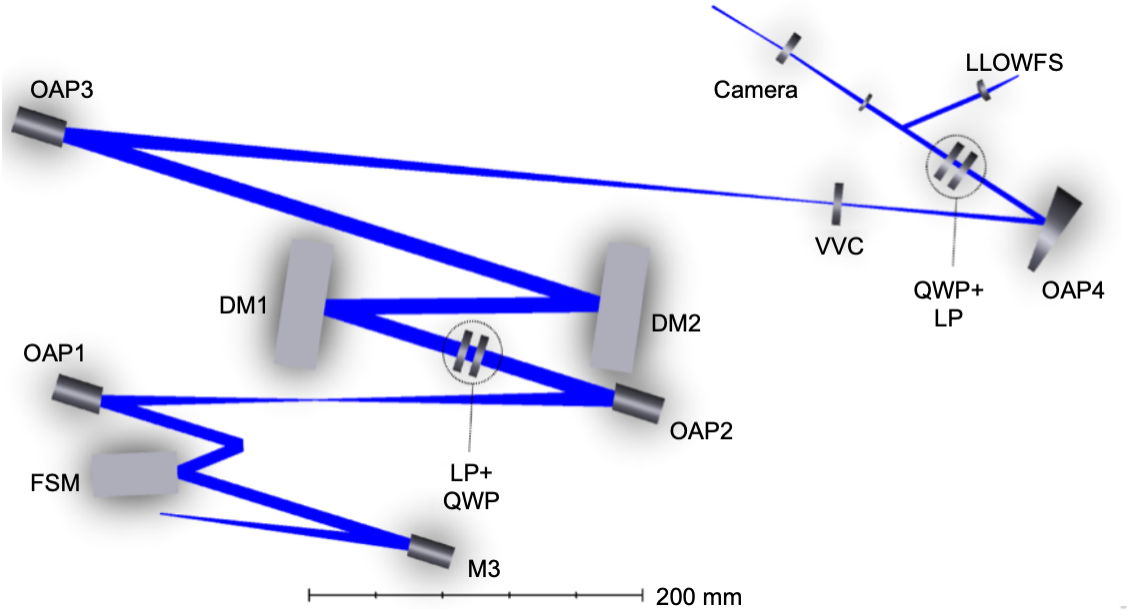}
    \caption{Raytrace of a candidate future layout for SCoOB including a second DM to correct for amplitude errors that arise from pupil discontinuities.}
    \label{fig:two_dms}
\end{figure}

The Space Coronagraph Optical Bench is a new high-contrast imaging instrument undergoing active development at the UA Space Astrophysics Laboratory. We present the modified design of the high contrast imaging instruments, the quality of the optics, and the materials and process by which the testbed is assembled.
In the immediate future the SCoOB team will focus on refining the contrast achievable out of vacuum and eliminating errant noise sources that could limit our testbed performance. Recently the SCC Lyot stop has been installed, and we will be conducting tests of dark hole stability under SCC control in the near future. To learn about our wavefront sensing and control efforts in detail, see the accompanying proceedings by Van Gorkom et al \cite{VanGorkom2022}. A candidate future realization of SCoOB includes two deformable mirrors (Fig \ref{fig:two_dms}) to correct for amplitude errors from pupil segmentation\cite{Mazoyer2017}. A list of the hardware used in this testbed is included in the appendix after the references.

\section{Acknowledgements}
Many thanks to the many teams building and simulating existing testbed who have shared their expertise and know-how. In particular, the authors would like to thank Iva Laginja, Dimitri Mawet, Chris Mendillo, Mamadou N'Diaye, Marshall Perrin, A.J. Riggs, Garreth Ruane, and Remí Soummer for particularly helpful conversations and tours.
Thanks to the HiCAT team for sharing their design for the BMC Kilo-DM's mounting hardware\cite{Choquet_2018}, and a very helpful set of proceedings that guided the design phase of the testbed when it was called CDEEP \cite{N_Diaye_2013,N_Diaye_2014}.
Portions of this work were supported by the Arizona Board of Regents Technology Research Initiative Fund (TRIF).

\section{Appendix: Hardware used in the SCoOB Testbed}
\begin{table}[H]
    \centering
    \begin{tabular}{|c|c|}
    	\hline
        Component & Vendor   \\
        \hline
        Deformable Mirror & \href{https://bostonmicromachines.com/products/deformable-mirrors/standard-deformable-mirrors/}{Boston Micromachines Kilo-C 1.5um DM} \\
        Science Camera (in air only) & \href{https://starizona.com/products/zwo-asi-294mm-pro?_pos=1&_sid=e1ce075ce&_ss=r}{ZWO ASI 294mm Pro}\\
        Pupil Camera & \href{https://www.edmundoptics.com/p/Basler-ace-acA2500-60um-Monochrome-USB-30-Camera/34884/}{Basler Ace AC2500} \\
        Off-Axis Parabolic Mirrors & Nu-Tek\\
        6DOF OAP Mounts & \href{https://www.thorlabs.com/thorproduct.cfm?partnumber=K6XS#ad-image-0}{THORLABS K6XS} \\
        High-contrast Linear Polarizers & \href{https://www.meadowlark.com/precision-linear-polarizer/}{Meadowlark Precision Linear Polarizer} \\
        Quarter-wave plates & \href{https://boldervision.com/waveplates/aqwp3/}{Boulder Vision Optik AQWP3}\\
        Passivated 1" Posts & \href{https://www.newport.com/p/9961}{Newport 9961 Pedastal} \\
        1" Optic Mounts & \href{https://www.newport.com/p/U100-A2H-V6}{Newport 1" Mirror Mount} \\
        Flat Mirrors & \href{https://www.newport.com/p/10Z40AL.2}{Newport Aluminum Mirror} \\
        Vacuum-Compatible Breadboard & \href{https://www.newport.com/p/SA2-30x30-LC}{Newport 30" x 30" plate} \\
        \hline 
    \end{tabular}
    % \caption{Caption}
    \label{tab:components}
\end{table}

% References
\bibliography{report} % bibliography data in report.bib
\bibliographystyle{spiebib} % makes bibtex use spiebib.bst
\end{document}